 \documentclass[]{emulateapj}
\usepackage{amsmath}	% Advanced maths commands
\usepackage{amssymb}	% Extra maths symbols
\usepackage{cases}
\usepackage{threeparttable}
\usepackage{chngcntr}
\usepackage{footmisc}
\usepackage[breaklinks,colorlinks,citecolor=blue,linkcolor=magenta]{hyperref}

\graphicspath{{./}{figures/}}

\shorttitle{A young stellar ``snake" with two dissolving cores}
\shortauthors{Hai-Jun Tian}
\newcommand {\Usun}{{U_{\!\odot}}}
\newcommand {\Wsun}{{W_{\!\odot}}}
\newcommand {\Vsun}{{V_{\!\odot}}}

\newcommand{\kms}{km$\rm s^{-1}$}
\newcommand{\masyr}{\,mas\,$\rm yr^{-1}$}

\newcommand{\mul}{$\mu_{l^*}$}
\newcommand{\mub}{$\mu_{b}$}

\begin{document}

\title{Discovery of a young stellar ``snake" with two dissolving cores in the solar neighborhood}

\author{Hai-Jun Tian} %[0000-0001-9289-0589]
\affil{China Three Gorges University, Yichang 443002, China}
\affil{Center for Astronomy and Space Sciences, China Three Gorges University, Yichang 443002, China}
%\affil{Max Planck Institute for Astronomy, K\"onigstuhl 17, D-69117 Heidelberg, Germany}

%\author[0000-0002-1802-6917]{ETC}
%\affiliation{Key Lab for Optical Astronomy, National Astronomical Observatories, Chinese Academy of Sciences, Beijing 100012, China}
%\collaboration{(LaTeX collaboration)}



\begin{abstract}
We report the discovery of a young (only 30-40\,Myr) snake-like structure (dubbed a stellar ``snake") in the solar neighborhood from {\it Gaia} DR2. The average distance of this structure is about 310\,pc from us. Both the length and width are over 200\,pc, but the thickness is only about 80\,pc. The ``snake" has one tail and two dissolving cores, which can be clearly distinguished in the 6D phase space. The whole structure includes thousands of members with a total mass of larger than 2000\,$M_{\odot}$ in an uniform population. The population is so young that it can not be well explained with the classical theory of tidal tails. We therefore suspect that the ``snake" is hierarchically primordial, rather than the result of dynamically tidal stripping, even if the ``snake" is probably expanding. The coherent 5D phase information and the ages suggest that the ``snake" was probably born in the same environment as the filamentary structure of Beccari et al.(2020). If so, the ``snake" could extend the sky region of the Vela OB2 association by a factor of $\sim 2$, and supplement the census of its coeval structures. This finding is useful to understand the history of the formation and evolution of the Vela OB2 complex. The age of the ``snake" well matches with that of the Gould Belt. In the sky region of our interest, we detect one new open cluster, which is named as Tian 1 in this work.
%It has no significant leading tail, just a long trailing tail. 
%stripping or dynamical
%perhaps that is why we can not find any strong evidence for the theory of mass segregation
%and provide some clues to connect the complex with the Galactic-scale structures, e.g., the Gould Belt

\end{abstract}


%% Sections are demarcated by \section and \subsection, respective

\section{Introduction} \label{sec:intro}
It has long been recognized that O and B stars are not distributed randomly on the sky \citep{Eddington1910, Kapteyn1914}, but instead are concentrated in spatially loose, gravitationally unbound, and kinematically associated groups. Since \citet{Ambartsumian1947} initially introduced the term ``association" for the groups of OB stars, the historical name OB association has been widely used in the literature. Besides bright O and B members, these groups actually include a number of lower-mass stars and pre-main-sequence (PMS) stars over large areas, and follow a continuous mass function \citep{Brice2007}. As ideal test-benches, the nearby OB associations have been well studied for many years to understand the fundamental questions, e.g.,  how young stellar populations form in molecular clouds (MCs), and how they leave their natal gas and disperse into the Galactic field \citep{Blaauw1964, Zeeuw1999, Brice2007b, Armstrong2018, CG2019a}.
%and \citet{Ruprecht1966} compiled a list of OB associations, 

There are at least two typical mechanisms to explain the origin of OB associations in the literature. One is that associations are the unbound remnants of previously bound clusters \citep{Tutukov1978}. In this monolithic model of star formation, it is generally understood that the vast majority of stars form in embedded clusters \citep{Lada2003}, observed low density associations must have been formed as single dense stellar clusters that subsequently underwent a period of expansion to form the large scale  and unbound structures caused by physical processes such as feedback from photoionizing radiation and stellar winds. This model has often been used to explain the low proportion of gas-free and gravitationally bound clusters after a few Myr \citep{Hills1980,Kroupa2001,GB2006}. Alternatively, in a hierarchical model of star formation \citep[e.g.][]{Elmegreen2002, Bonnell2011}, stars form over substructured regions with various densities in MCs. Gravitationally bound stellar clusters arise naturally at high local density regions, while unbound and hierarchically-structured associations form in-situ at low density regions \citep{Kruijssen2012}. This model can naturally explain the wide range of observed stellar densities.  % when the residual gas is expelled

There has long been debate about whether the system forms in the monolithic or hierarchical models \citep{Muench2008}. \citet{Bressert2010} investigated the spatial distribution of the young stellar objects (YSOs) in the solar neighborhood and found that the nearby YSOs follow a smooth surface density distribution without features of multiple discrete modes. This is likely a result of star formation occurring over a continuous density distribution. However, this has been argued by studies showing that expanding clusters can reproduce a similar result \citep{Gieles2012, Pfalzner2012}. Significantly, some recent works have begun to challenge the first model. \citet{WK2018} quantified the kinematics of 18 nearby OB associations in {\it Gaia} TGAS and conclusively ruled out the monolithic cluster formation scenarios. \citet{WM2018} analyzed the kinematics of the Scorpius-Centaurus OB association from {\it Gaia} DR1 \citep{gaia2016} and did not find the coherent evidence for the classical theory that OB associations are the expanded remnants of dense and compact star clusters.  More recently, nevertheless, \citet{Kuhn2019} investigated the kinematics of 28 young star clusters and associations in {\it Gaia} DR2 and revealed that at least 75\% of these systems are expanding with a median velocity of $\sim$0.5 \kms. It suggests that some young clusters still exhibit the potential to expand into large scale associations.
 
%This is a monolithic model of star formation. 
%Only the most massive (total mass $>$500\,\Msun) embedded clusters survive emergence from MCs to become bund systems, e.g., open clusters.

%After a few Myrs, newly formed O and B type stars sweep away the molecular cloud in which the cluster is embedded via feedback from photoionizing radiation and stellar winds, and the loss of this binding mass causes the cluster to become unbound and disperse.

%Members are too loosely populated to be gravitationally bound in the groups (Bok 1934; Mineur 1939).  their velocity dispersions are only a few kilometers per second (e.g., Mathieu 1986; Tian et al. 1996), and so they form coherent structures in velocity space.

%Vela OB2 complex is one of such test-benches, since it is the closest regions in the early stages of star formation at ages 10 to 50\,Myr,  and with a wide variety of different star formation environments. 

The Vela OB2 is one of closest associations spanning a wide sky region near the border between the Vela and Puppis constellations. Since it was first reported by \citet{Kapteyn1914}, Vela OB2 has been known for decades as a sparse group of a dozen early-type stars \citep{Blaauw1964, Brandt1971, Straka1973}. \citet{Zeeuw1999} pioneeringly investigated its kinematical features based on the the astrometry of Hipparcos, and noted that the space motion of the association is not clearly separate from the surrounding stars and the nearby open clusters NGC 2547 and Trumpler 10. \citet{Pozzo2000} reported a population of low-mass and PMS stellar association (often referred to as $\gamma$ Vel cluster) nearby its brightest binary member $\gamma^2$ Vel in a distance of 350-400pc from us. \citet{Jeffries2009} claimed that the $\gamma$ Vel cluster is a subcluster within the larger Vela OB2 association, and speculated that $\gamma^2$ Vel (age 3-4\,Myr) formed after the bulk of the low-mass stars (age $\sim$10\,Myr), expelling gas, terminating star formation and unbinding the association. Making use of data from the Gaia-ESO Survey \citep{Randich2013}, \citet{Jeffries2014} detected two kinematic sub-groups from the $\gamma$ Vel cluster, and suggested that one group (Population A) might be related to $\gamma^2$ Vel. Subsequently, Sacco et al. (2015) identified that the radial velocities in another group (Population B) are consistent with that of the nearby cluster NGC 2547 (age $\sim$35$\pm$3\,Myr from \citet{Jeffries2005}; distance 361$^{+12}_{-11}$\,pc from \citet{Jeffries2009}). The bimodal feature is confirmed by \citet{Damiani2017} with TGAS data in the 2D proper-motion space. \citet{Franciosini2018} used the {\it Gaia} DR2 data to resolve the 6D structure of the  $\gamma$ Vel cluster, and find that both of the nearly coeval populations differ not only kinematically, but are also located at different distances along the line of sight. Furthermore, \citet{Beccari2018} noted that the region might host not two but six kinematic groups, four of which are coeval with the $\gamma$ Vel cluster, while the remaining two probably formed together with NGC 2547. Most recently, \citet{CG2019a, CG2019b} identified at least seven groups from the Vela OB2 complex in the 6D phase space with the {\it Gaia} DR2 data, and confirmed that their overall structure is expanding, which suggests a common history for Vela OB2 and the IRAS Vela Shell \citep[IVS,][]{Sahu1992}. The current kinematics of the stars can not be explained by internal processes alone, e.g., gas expulsion.

In this work we report a young and wide (over 200\,pc) structure with two dissolving cores in the solar neighborhood from the data of {\it Gaia} DR2. The structure is so young (only 30$\sim$40\,Myr) that it can not be well explained with the classical theory of tidal tails. It may primordially form from a huge MC. Conservatively, we name it a stellar ``snake" by its morphology in the sky plane. This morphology looks like the string-like populations reported by \citet{Kounkel2019} and \citet{Kounkel2020}, who identified thousands of such structures. Interestingly, the ``snake" probably is a good extension to the census of coeval structures in the Vela OB2 complex, since its age is well consistent with the 260\,pc wide filamentary population uncovered by \citep[][hereafter B20]{Beccari2020}, which bridges the NGC 2547 and the newly detected cluster BBJ 1.
%In this study we measure the detailed features including the age, mass, spacial and kinematical distributions 

%Perhaps it is not a canonical tidal tail, and may not even be a tidal tail.
%Its age is only 30$\sim$40\,Myr, size is over 200\,pc, and total mass is more than 2000 $M_{\odot}$. This finding makes a great challenge to the theory of tidal tail formation and evolution from a star cluster \citep{Kharchenko2009}, 

%and probably fill up the observational gap between the Gould Belt \citep{poppel1997} and star formation nearby the Orion complex\footnote{The Orion complex is thought to form part of the Gould Belt, which is a large ($\sim$1\,kpc), young (30 - 40\,Myr), and ring-like structure tilted $\sim20\degr$ to the Galactic plane. But the oldest stellar population observed so far is only 21\,Myr \citep{Kos2019}. The stellar population discovered in this work is well consistent with the history of Gould Belt.}.

%According to the high resolution N-body simulations carried out by Kharchenko et al. (2009)

The main goal of this study is to depict the detailed features for the ``snake" structure with the data of {\it Gaia} DR2, and to look for some clues to its formation and evolution. To do so, we organize this paper as follows. In Section \ref{sec:data}, the sample selection and membership are briefly described. The population properties are derived in Section \ref{sec:results}. The discussion and conclusion parts are in Section \ref{sec:dis} and \ref{sec:conclusion}, respectively.

Throughout the paper, we adopt the solar motion as $(\Usun,\Vsun,\Wsun) = (9.58, 10.52, 7.01)$\kms\, \citep{tian2015} with respect to the local standard of rest (LSR) , and the solar Galactocentric radius and vertical positions as ($R_0$, $z_0$) = (8.27,0.0)\,kpc \citep{schonrich2012}. $l^{*}$ is used to denote the Galactic longitude in the gnomonic projection coordinate system, for example,  $\mu_{l^{*}}\equiv \mu_{l}\cos b$. The proper motion ($\mu_{l^{*}}, \mu_b$) for each star is calibrated from the effect of the Solar peculiar motion in the Galactic coordinate.

%the circular speed of the as $v_0=238$\kms \citep{schonrich2012}

\section{Data and membership} \label{sec:data}
In this study, we mainly use the astrometric and photometric data from {\it Gaia} DR2 \citep{gaia2018} to search for members of the intended structure.

% by applying two hierarchical clustering algorithms
%5-dimensional astrometric data
%and color magnitude diagram (CMD)

\subsection{Data selection} \label{sec:data_selection}
%We initially noticed the structures in the proper motion distribution for the stars with an typical distance of $\sim$305\,pc in the sky region nearby Orion complex. There is a significant feature of overdensity centered at (-2.59, -2.0) \masyr in the space of (\mul, \mub). 
To build the sample, we select the sources that satisfy the following criteria:
\begin{enumerate}
 \item $170{\degr}<l<225{\degr}$, and $-30{\degr}<b<10{\degr}$, to make sure all the scattered members are included in this sky region. 
 %Note that the selected sky region is larger than that adopted in the other literatures, since we preliminarily found the members almost sparsely fill up the sky region of $\lambda$ Orion.
 \item $2.0<\varpi<5.0$\,mas, and $\varpi/\sigma_{\varpi}>10.0$, to restrict the sample to sources with distances between 200\,pc and 500\,pc. Here, we derive distances by inverting parallax with $d=1000.0/\varpi$\,pc.
 \item $(\mu_{l^*} - \bar{\mu}_{l^*,m})^2 + (\mu_{b} - \bar{\mu}_{b, m})^2<(5\sigma_{\mu, m})^2$, to restrict to stars with proper motions within 5$\sigma_{\mu, m}$ of ($\bar{\mu}_{l^*, m}$, $\bar{\mu}_{b, m}$). Here, $\bar{\mu}_{l^*, m}$, $\bar{\mu}_{b, m}$, and $\sigma_{\mu, m}$ are the average and root-mean-square ({\it rms}) of proper motions of the members in the stellar ``snake". We noticed the structure in the space of (\mul, \mub) initially as an overdensity centered on ($\bar{\mu}_{l^*, m}$, $\bar{\mu}_{b, m}$) $\simeq (-2.59, -2.0)$ \masyr\ with a dispersion $\sigma_{\mu, m} \simeq 1.0$ \masyr.
 \item $\rm RUWE<1.4$, to limit to sources with acceptable astrometric solutions. $\rm RUWE$ is the renormalized unit weight error which is defined in \citet{Lindegren2018} 
 %\begin{equation}
 %\sqrt{\chi^2/(\nu'-5)}/ \max(1, \exp(-0.2(G-19.5)), 
 %\end{equation}
 %where $\chi^2$ and $\nu'$ are respectively referred to as the astrometric goodness-of-fit  and the number of good observations in the AL direction \citep{Lindegren2018}. % (\texttt{astrometric\_chi2\_al})(\texttt{astrometric\_n\_good\_obs\_al})
\end{enumerate}
These selection criteria yield a sample of 21,949 stars in total. The interstellar extinction has been corrected for each source in the sample. According to the literatures \citep{MA1980, Wang2017}, the interstellar extinction in the V band is around 0.7-1.0\,mag/kpc in the solar neighborhood. For simplicity, an average value (i.e., 0.85\,mag/kpc) is adopted in this study \citep{LJM2020}. So for each individual star at distance $d$, its V-band extinction $A_V$ in magnitude is approximately $0.85 \times d$. The extinctions in {\it Gaia}'s bands for each star can be calculated from $A_V$ \citep{tian2014}. \citet{WC2019} provides the extinction coefficients ($A_{\lambda}/A_V$) for {\it Gaia}'s three bands, which are 1.002, 0.589, and 0.789 $\mu m$ for the $G_{BP}$, $G_{RP}$, and $G$ bands, respectively. 

\begin{figure*}[!t]
\centering
\includegraphics[width=0.85\textwidth, trim=0.0cm 0.0cm 0.cm 0.0cm, clip]{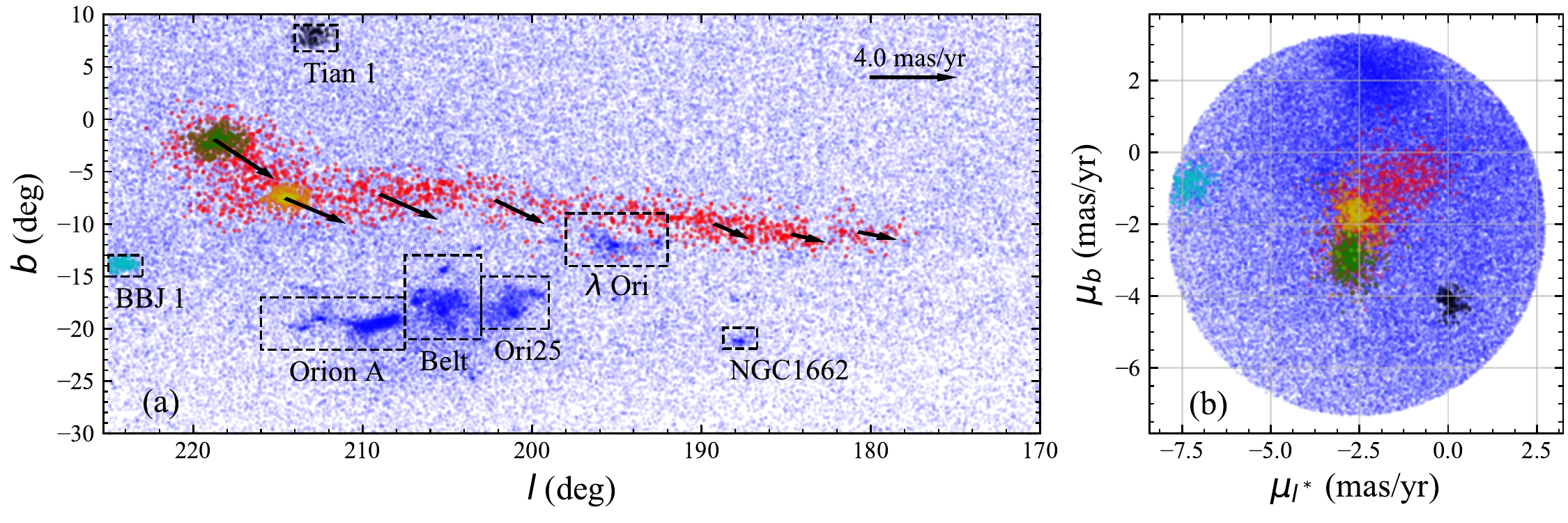}
\includegraphics[width=0.85\textwidth, trim=0.0cm 0.0cm 0.cm 0.0cm, clip]{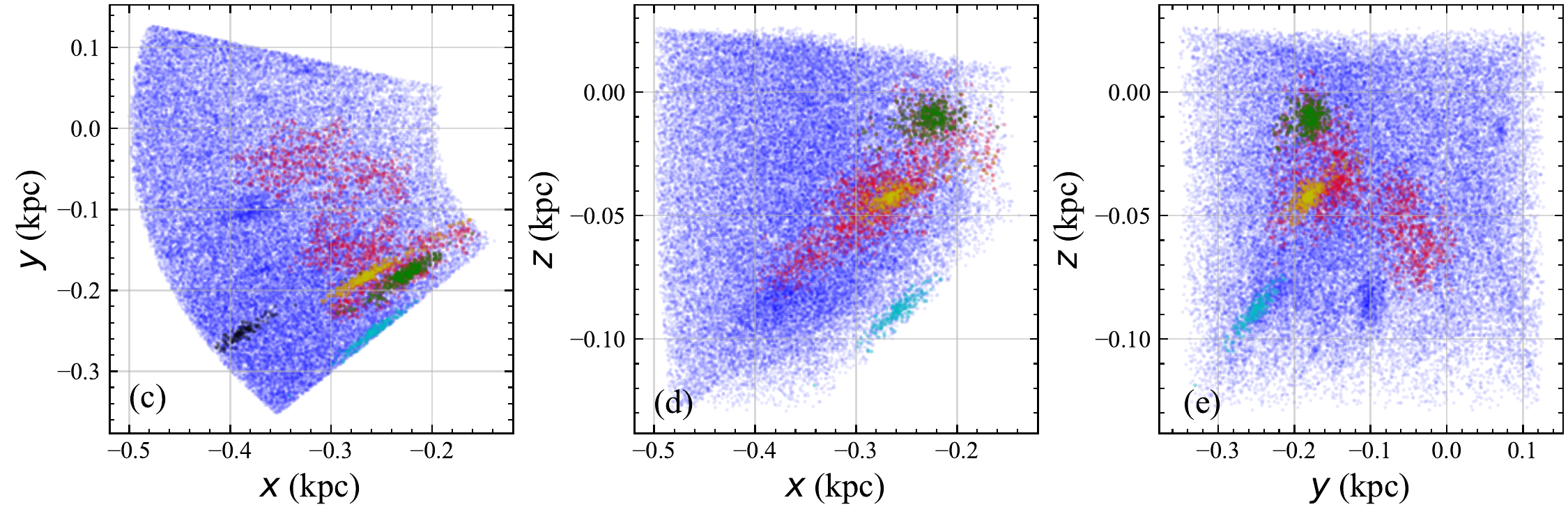}
\caption{The sample distribution in the 5D phase space, e.g., the $l$-$b$ projected space in the sub-panel (a), the proper motion (\mul-\mub) space in the sub-panel (b), and the 3D (x-y-z) spatial spaces in the sub-panels (c)-(e). In all the sub-panels, the red dots represent the candidate members of the stellar ``snake", the green and yellow dots show the two dissolving cores (the yellow core is actually known open cluster NGC 2232), while the cyan and black dots represent the open cluster BBJ 1 (B20) and an unknown cluster (which is temporarily named as Tian 1 in this study)}. In the sub-panel (a), the black arrows demonstrate the average proper motions in different $l$ bins observed in the frame of the LSR.  The black dashed rectangles label the known structures, such as $\lambda$ Ori, 25 Ori, Belt, Orion A, NGC 1662, and BBJ 1. Note that: (1) the proper motion direction vary smoothly with $l$,  and the amplitude becomes smaller towards the end of its tail with vanishing proper motion in the $b$ direction at the tail; (2) the stellar ``snake" is still clearly visible only on the members with $M_G<9.0$\,mag; (3) the cluster Tian 1 (the black dots) is not displayed in the x-z and y-z sub-panels, because its coordinate is about 0.1\,kpc in the z-direction, which is out of the displaying region in the sub-panels (d) and (e).
\label{fig:data_distr}
\end{figure*}

Figure \ref{fig:data_distr} displays the sample distributions in the 5 dimensional (5D) phase space. E.g., the $l$-$b$ projected space (sub-panel a)
%, the 3D (x-y-z) spaces in the sub-panels (c)-(e), and the proper motion (\mul-\mub) space in the sub-panel (b). 
shows there are several star forming regions (black dashed rectangles) in this field, such as the Orion complex \citep{Zari2019, Chen2019}, which is grouped into several clustering components, e.g., $\lambda$ Ori, 25 Ori, Belt, Orion A, BBJ 1, and so on.

\subsection{Membership} \label{sec:data_selection}
Considering that members of the structure are loosely concentrated in the 3D space, we adopt the ``Friend-of-Friend" (``FOF") algorithm to search for members. This algorithm has been well implemented in ``ROCKSTAR" \citep{Behroozi2013}, which employs a technique of adaptive hierarchical refinement in the 6D phase space to divide all stars into several ``FOF" groups by tracking the high number density clusters and excluding those stars that could not be grouped in star aggregates.

The majority of sources in our sample has only 5D phase information, i.e., $l$, $b$, \mul, \mub, and distances, obtained from {\it Gaia} DR2. Line-of-sight velocities ($v_{los}$) are provided only for stars brighter than 13\,mag in the G-band. However, the inputs of ROCKSTAR are originally designed to be 6D phase-space data sets. We test  on a version of ROCKSTAR which was smartly optimized by \citet{TianH2017}, and find that ROCKSTAR works well for group searching if we set $v_{los}$ as zero, and keep only the other 5D data for all sample stars. In the process, ROCKSTAR will automatically adjust the linking space between members of ``friend" stars, and divide them into several groups, simultaneously single out those isolated individual stars. In this step, we get 2175 candidate members of the stellar ``snake" from the input sample. This process is valid as we can recover the members of all the known clusters in this region simultaneously, i.e., $\lambda$ Ori, 25 Ori, Belt, Orion A, and NGC 1662 from the FOF groups. This indicates that ROCKSTAR works very well for group searching.

Since the members are pretty sparse in space, and the proper motions are not uniform along the long tail, particularly close to the end of the tail, it is a challenge to obtain candidate members with a high fidelity for any clustering algorithm. Some very young candidates obviously deviate from PMS or main sequence (MS) in the Colour-Absolute Magnitude Diagram (CAMD). To eliminate this contamination, we remove candidates whose ages are beyond 120\,Myr or younger than 5\,Myr (see the green dashed and dashed-dot curves in the left sub-panel of Figure \ref{fig:iso_fitting}). In this step, 186 stars are rejected. Finally, we obtain 1989 candidate members in the tidal tail. 
%Actually only 1950 stars are used for this study, because 81 candidates are ruled out according to their radial velocities (see Section \ref{sec:z_rv}).

Figure \ref{fig:data_distr} displays the member (the red dots) distributions in the 5D phase space. A structure looking like a snake is clearly demonstrated in the ordinary space (i.e., $\alpha$-$\delta$, x-y-z). %Its original open cluster can be found from the panels of $l$-$b$ and \mul-\mub. 
% ($\alpha$-$\delta$, x-y-z), \mul-\mub\ 

\section{Population Property} \label{sec:results}
We have 1989 candidate members which can be used to analyze the statistical properties for the stellar ``snake". Besides the astrometric and photometric data from {\it Gaia} DR2,  we cross-match the sample with the spectroscopic data from LAMOST \citep{zhao2012, cui2012} and APOGEE \citep{Abolfathi2018} to supplement  the measured stellar parameters of members. Within a radius of 3 arcseconds, 78 and 15 candidate members are matched with LAMOST DR6 and APOGEE-2, respectively.

\subsection{Spatial distribution and steller parameters}\label{sec:z_rv}
Figure \ref{fig:data_distr} shows the member distributions in the 5D phase space. The red points in the space of $l$-$b$ (sub-panel (a)) clearly demonstrate the shape of the tidal tail. Both the length (in the x-direction) and width (in the y-direction) are over 200 pc, but the thickness (in the z-direction) is only around 80\,pc, as shown in the sub-panels (c-e). The average proper motion (\mul, \mub) is about (-2.50$\pm$0.68, -1.84$\pm$1.07) \masyr\ (see the sub-panel (b)), and the values become smaller towards the end of its tail. Meanwhile, the proper motion gradually vanishes in the $b$ direction, as shown by the black arrows in the sub-panel (a). This should be due to the variation of the distances from us in the different parts of the ``snake".

Among the candidates, only 243 stars have radial velocities, of which 176 stars from {\it Gaia}, 78 stars from LAMOST (23 stars also have  {\it Gaia}'s radial velocities), and 15 from APOGEE (3 stars also have  {\it Gaia}'s radial velocities), and 15 from APOGEE). Figure \ref{fig:hist_dist_rv} presents the histogram of the radial velocities of 243 unique stars in the left panel. The average and root-mean-square ({\it rms}) of the radial velocities are about 24.7 and 4.7 \kms. Figure \ref{fig:hist_dist_rv} also shows the histogram of the distances of the 1989 candidate members in the left panel. The average and {\it rms} of the distances are around 310.4 and 39.6\,pc, respectively. The {\it rms} value is larger than what one expects, as there exist two cores separated slightly in the stellar ``snake". We postpone its discussion to the next subsection.

The element abundance and $\log$g can be partly obtained from the LAMOST and APOGEE data. Table \ref{tab:stellar_para} specifies the stellar parameters for the candidate members which have effective values of [Fe/H] and precise values of $r_v$. The average and {\it rms} of [Fe/H] are 0.019$\pm$0.08 for all the 40 available candidates, and 0.028$\pm$0.08 for the 28 members whose memberships are further confirmed by their radial velocities, e.g., $18<v_r<28$\,kms. Therefore, a super-solar [Fe/H] is favored in this data set. The average and {\it rms} of $\log$g are 4.31$\pm$0.18 for all the available candidates. The stellar parameters for the 1989 candidates are available in the electronic version.

%\clearpage
%\begin{landscape}
\begin{table*}
\begin{threeparttable}
\caption{ The stellar parameters of 10 members with metallicities and $\epsilon_{r_v}<3$\,\kms}.\label{tab:stellar_para}
\centering
\begin{tabular}{c|c|c|c|c|c|c|c|c|c|c}
\hline
\hline
{\it Gaia} ID&$l$\footnotemark&$b$&\mul\footnotemark&\mub&Distance&$v_r$&[Fe/H]&$\log$g&$G$&BP-RP\\
\hline
&\multicolumn{2}{c|}{\degr(J2000)}&\multicolumn{2}{c|}{\masyr}&pc&\kms&&&\multicolumn{2}{c}{mag}\\
\hline
3104206755858178816&214.81&-7.79&-2.70$\pm$0.04&-1.69$\pm$0.04&327.61$\pm$2.44&25.77$\pm$0.04&0.04$\pm$0.01& &5.83&1.21\\
3118540642272975872&210.39&-7.80&-2.59$\pm$0.05&-1.35$\pm$0.05&304.22$\pm$2.84&25.62$\pm$0.04&0.08$\pm$0.01&4.41$\pm$0.29&6.88&1.63\\
3118203882476896768&211.32&-7.59&-2.41$\pm$0.06&-1.44$\pm$0.05&306.67$\pm$2.93&25.18$\pm$0.03&-0.01$\pm$0.01& &6.36&1.39\\
3121980223881083264&208.91&-7.66&-2.51$\pm$0.06&-1.22$\pm$0.06&293.89$\pm$4.02&24.68$\pm$0.04&0.14$\pm$0.01& &7.23&1.79\\
3409223204229856640&180.58&-11.75&-0.89$\pm$0.07&0.02$\pm$0.04&334.11$\pm$4.12&21.78$\pm$0.02&0.13$\pm$0.01& &6.59&1.52\\
3418236450799000320&180.36&-11.50&-1.73$\pm$0.09&-0.04$\pm$0.06&288.61$\pm$3.71&20.78$\pm$0.03&0.06$\pm$0.01& &7.14&1.71\\
3394383782983956864&186.61&-12.21&-0.56$\pm$0.09&-0.02$\pm$0.07&370.87$\pm$5.79&24.80$\pm$0.03&0.05$\pm$0.01& &6.45&1.43\\
3389597883746236800&191.07&-11.14&-2.37$\pm$0.06&-0.36$\pm$0.05&267.96$\pm$2.61&22.71$\pm$0.02&0.08$\pm$0.01& &7.16&1.74\\
3394543177810189312&186.18&-11.35&-0.43$\pm$0.07&-0.15$\pm$0.05&385.07$\pm$5.31&25.57$\pm$0.04&0.03$\pm$0.01& &6.20&1.41\\
3336872762142253824&196.00&-8.16&-2.69$\pm$0.05&-1.18$\pm$0.04&248.25$\pm$1.89&24.87$\pm$2.48&-0.02$\pm$0.08&4.50$\pm$0.14&6.38&1.42\\
\hline
\hline
\end{tabular}
\begin{tablenotes}
  \item [a] The on-line coordinate for each star will be given in the precision of double float.
  \item [b] The effect of the peculiar motion of the Sun are removed from the proper motions. 
 % \item [c] The radial velocities are from {\it Gaia} DR2, LAMOST DR6, and APOGEE-2.
 % \item [d] The metallicities are from LAMOST DR6, and APOGEE-2. 
  %\item [e] All the $\log$g are obtained only from LAMOST DR6. 
  \end{tablenotes}
 \end{threeparttable}
\end{table*}
%(the full candidates are available on-line)
%\end{landscape}
%\clearpage

\subsection{The double dissolving cores}
\begin{figure}[!t]
\centering
\includegraphics[width=0.245\textwidth, trim=0.22cm 0.cm 0.cm 0.cm, clip]{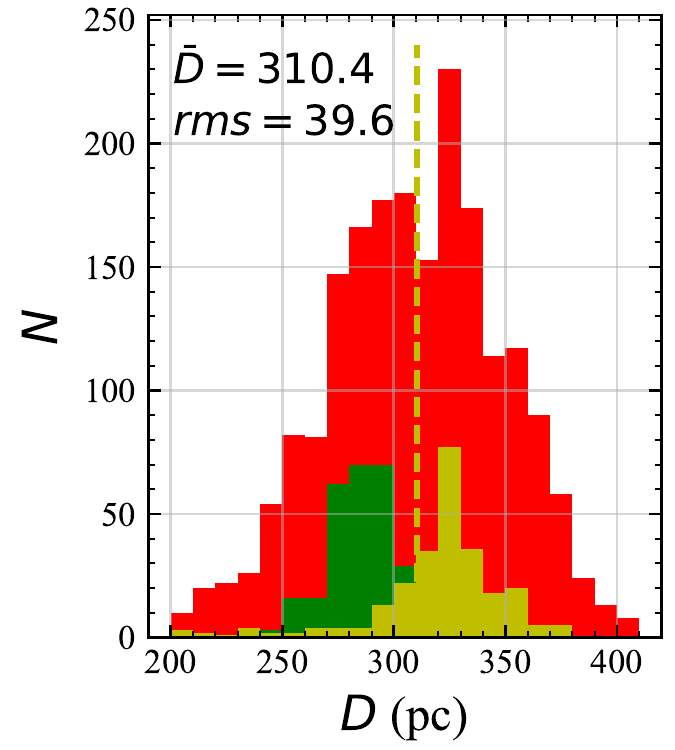}
\includegraphics[width=0.22\textwidth, trim=0.85cm 0.cm 0.cm 0.cm, clip]{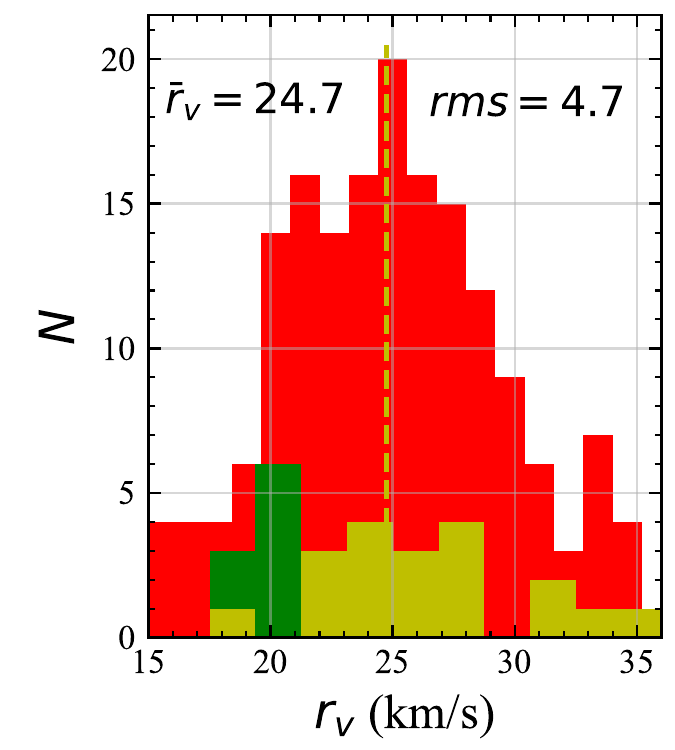}
\includegraphics[width=0.245\textwidth, trim=0.22cm 0.cm 0.cm 0.cm, clip]{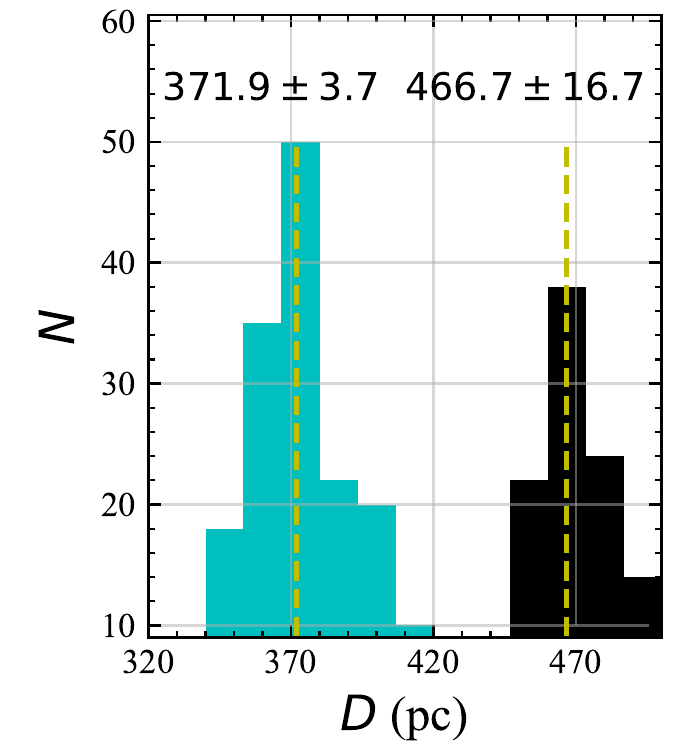}
\includegraphics[width=0.22\textwidth, trim=0.85cm 0.cm 0.cm 0.cm, clip]{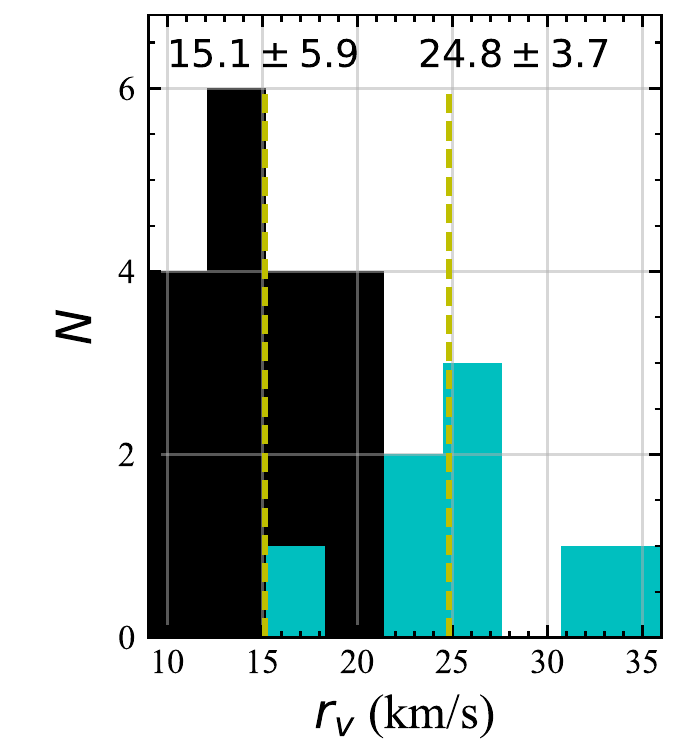}
\caption{ Histograms of the distances and radial velocities of the whole candidates (red), the two dissolving cores (green and yellow), the open cluster BBJ 1 (cyan) and Tian 1 (black). The average distances are 310.4$\pm$ 39.6\,pc, 288.95$\pm$16.70\,pc, 324.17$\pm$18.32\,pc, 371.9$\pm$3.7\,pc, and 466.7$\pm$16.7\,pc for the whole candidates, the two cores, and the open clusters BBJ 1 and Tian 1. The corresponding radial velocities are 24.8$\pm$4.5\,\kms, $22.3\pm3.23$\,\kms, $26.0\pm3.5$\,\kms, 15.1$\pm$5.9\,\kms, and 24.8$\pm$3.7\,\kms, respectively. The yellow dashed vertical lines mark the average distances and radial velocities written on the sub-panels.
}\label{fig:hist_dist_rv}
\end{figure}

The most prominent feature is that both the radial velocities and distances present two explicit peaks in their histograms (see the top two sub-panels of Figure \ref{fig:hist_dist_rv}). This is the result of two cores (or clusters) in the structure of stellar ``snake". In order to explicitly display the two cores, we select two sub-samples according to their coordinates, e.g., $(l - l_0)^2 + (b - b_0)^2<\epsilon_0^2$, where $l_0$, $b_0$, and $\epsilon_0$ are the coordinate of the center of the core, and the angular radius of the cluster. We roughly estimate ($l_0, b_0,\epsilon_0$) = ($218.7\degr, -2.0\degr, 2.0\degr$) for the first core, and ($l_0, b_0,\epsilon_0$) = ($214.5\degr, -7.6\degr, 1.5\degr$) for the second core (this core actually is the known open cluster NGC 2232). Thus, we obtain 323 and 261 stars from the sample of candidate members. They are separately color-coded with green and yellow in Figures \ref{fig:data_distr} and \ref{fig:hist_dist_rv}. As one can see, the two cores are clearly separated in the 6D phase space. 

The average proper motions (\mul, \mub) are about (-2.75$\pm$0.27, -2.91$\pm$0.40) \masyr, and (-2.62$\pm$0.33, -1.76$\pm$0.37) \masyr\ for the two cores, respectively (see the green and yellow dots in the sub-panel (b) of Figure \ref{fig:data_distr}). The difference of the proper motions is larger than 1.0 \masyr\ between the two cores in the direction of Galactic latitude. The average distances are 288.95$\pm$16.70\,pc and 324.17$\pm$18.32\,pc for the two cores, respectively (the green and yellow histograms in the top-left sub-panel of Figure \ref{fig:hist_dist_rv}). The difference is about 35\,pc. While the available radial velocities have only 19 stars for both  cores, the $v_r$ of the two clusters can still be separated in their histograms (see the top-right sub-panel of Figure \ref{fig:hist_dist_rv}). The average values of $v_r$ are $22.3\pm3.23$\kms\ and $26.0\pm3.5$\kms. The difference is of a few \kms.

\begin{figure*}[!t]
\centering
\includegraphics[width=0.483\textwidth, trim=0.4cm 0.27cm 0.26cm 0.2cm, clip]{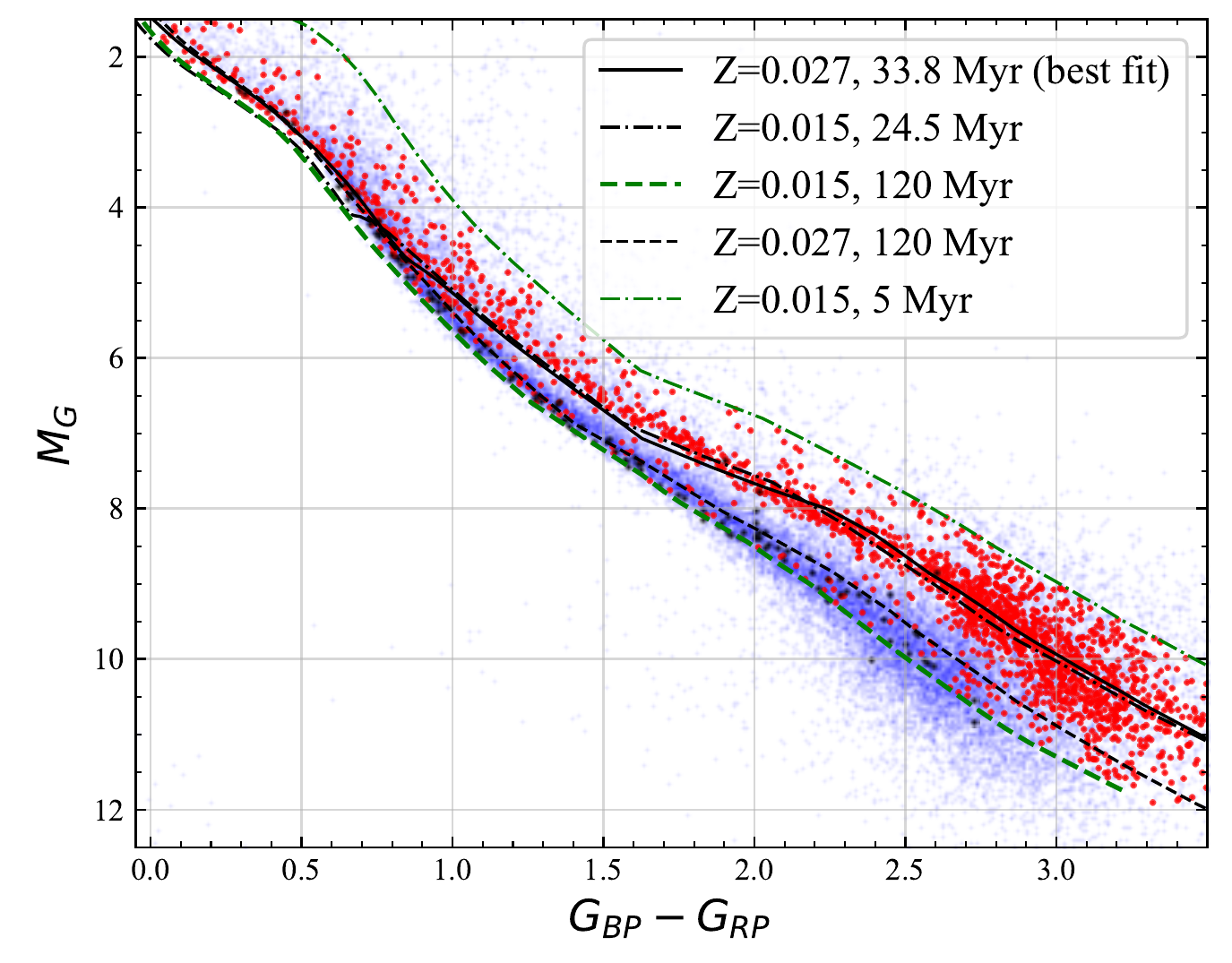}
\includegraphics[width=0.443\textwidth, trim=1.5cm 0.27cm 0.26cm 0.2cm, clip]{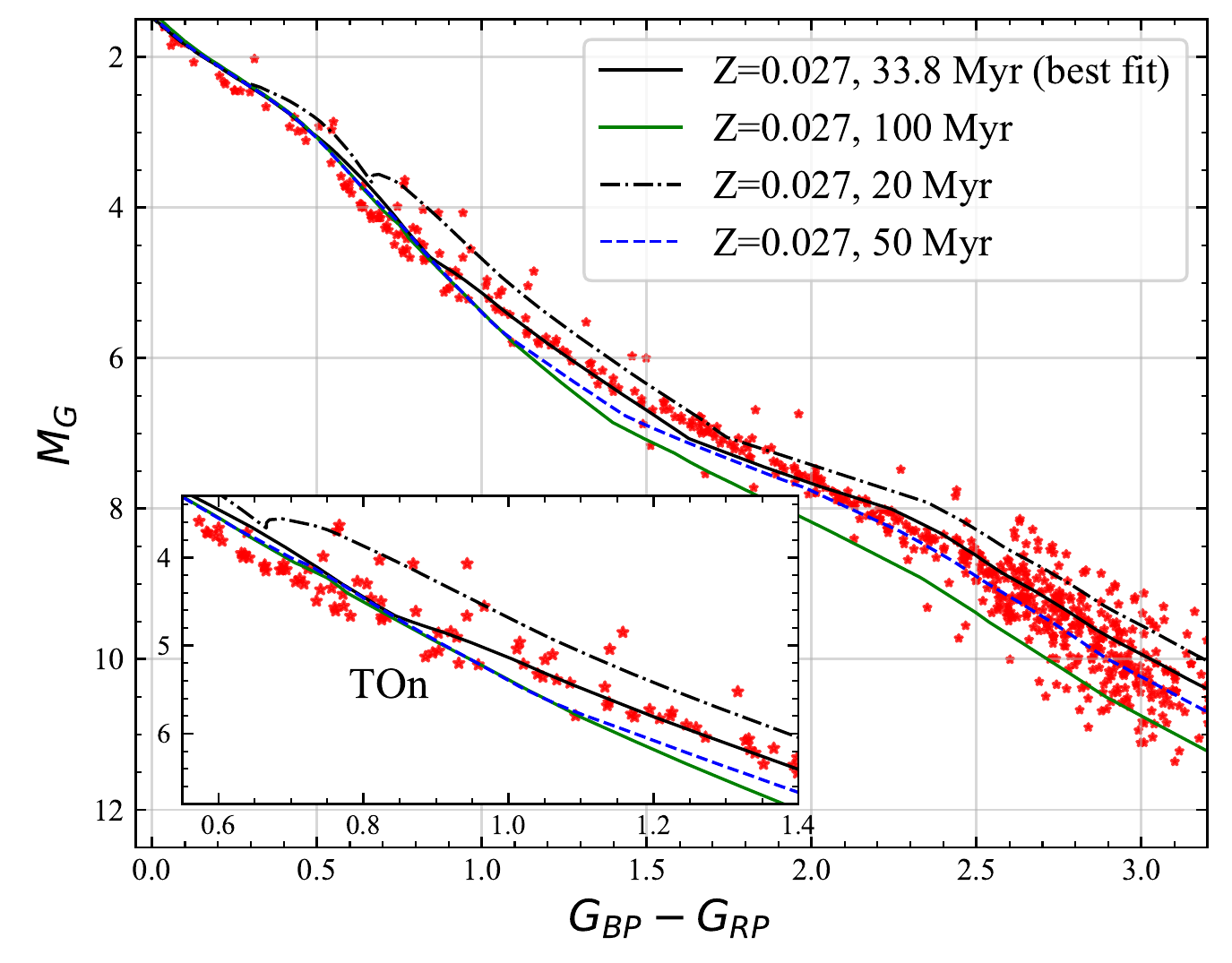}
\caption{Left: CAMD for the 1989 candidate members (the red dots), initial sample (the blue dots), and the open cluster Tian 1 (the black dots). The black-dashed curve is the best fitted isochrone from a subsample with $G_{BP}-G_{RP}<3$. The green dashed and dashed-dot curves are two isochrones to remove contaminations with ages $>$120\,Myr or $<$ 5\,Myr from the candidate members.   Right: CAMD for 634 members (red dots), which is a sub-sample purified with more stringent criteria from the 1989 candidates. In order to present the TOn point, we compare the best fit isochrone with a series of isochrones with different ages, e.g., 20\,Myr (the black dashed-dot curve), 50\,Myr (the blue dashed curve), and 100\,Myr (the green solid curve). The TOn point is clearly shown in the insert sub-panel. In both panels, the black solid curves represent the best fit PARSEC isochrone, the corresponding age is 33.8\,Myr, and metallicity (Z) is 0.027. 
}\label{fig:iso_fitting}
\end{figure*}

\begin{figure}[!t]
\centering
\includegraphics[width=0.4\textwidth, trim=0.1cm 0.07cm 0.cm 0.cm, clip]{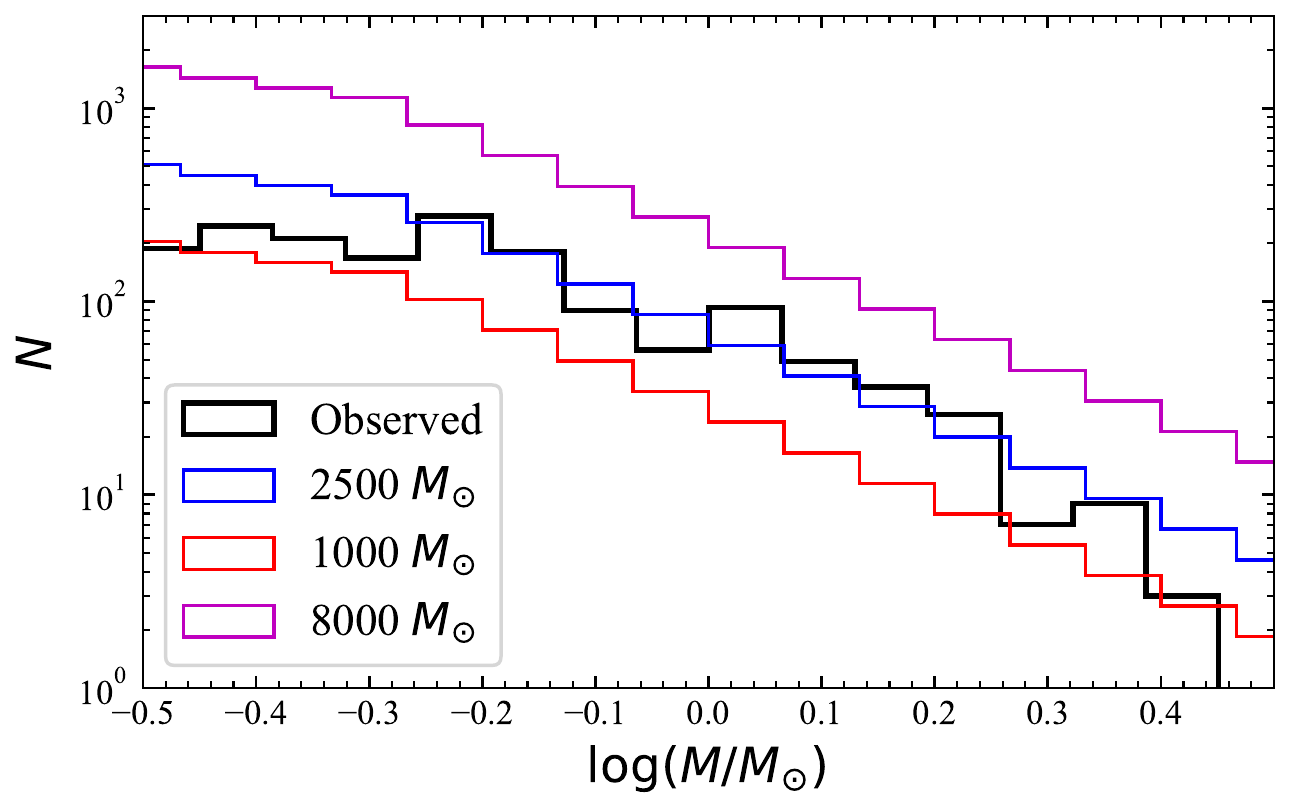}
\includegraphics[width=0.4\textwidth, trim=0.1cm 0.07cm 0.cm 0.cm, clip]{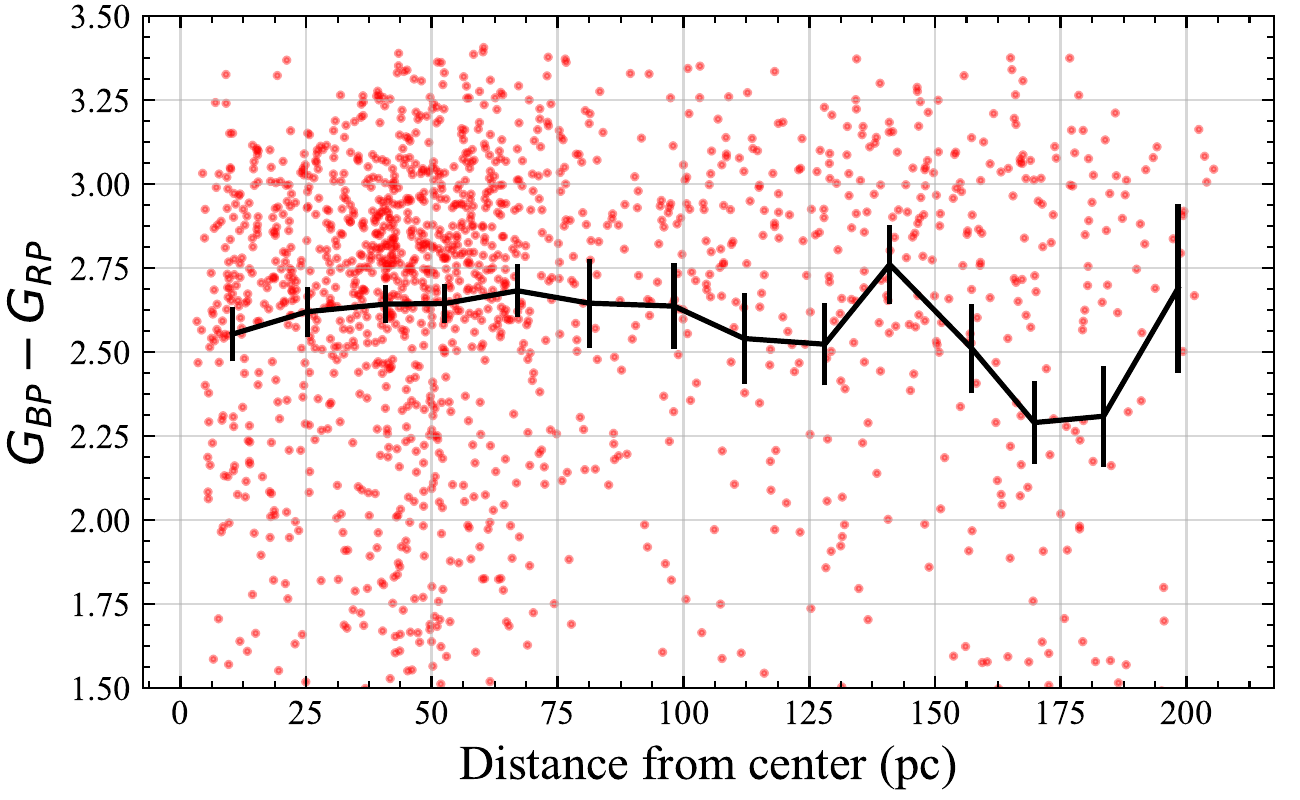}
\caption{ Top: Mass estimation of the stellar ``snake" based on the IMF model of \citet{Kroupa2001}. The black curve is the observed mass function, which is fitted with the IMFs with different total mass (displayed with different colors). As shown, the blue curve (2500 $M_{\odot}$) well matches the observed function. Bottom: Color index v.s. distance from the cluster center, to check the theory of mass segregation \citep{BT1987}. Here, the color is interpreted as a proxy of stellar mass. The cluster is centered at (x, y, z) = (-261, -175, -37.8)\,pc, which is the median position of the two cores. The black solid curve and the error are the average and uncertainty of the color in each bin. No strong evidence for mass segregation is found within 125\,pc from the cluster center. Note that the tendency of first rising then descending between 10 and 110\,pc is similar with what was found in \citet{MA2019}.
}\label{fig:mass}
\end{figure}

\subsection{Age and Mass}\label{sec:age}
Isochrone fitting is a typical method to estimate stellar ages. \citet{LP2019} provides a reliable implementation for this method. First, we prepare a series of isochrones from the Padova database \cite{Marigo2017} with ages ranging from $\log(\tau/yr)=6.6$ to 10.13 with an interval of $\Delta\log(t/yr)=0.01$, and metallicities from $\log(Z/Z_{\odot})=-2.0$ to 0.5 with an interval of 0.25. Then, we feed the color index ($G_{BP}$-$G_{RP}$), and the absolute magnitude $M_G$ of the member stars to the pipeline. Finally, the best fit isochrone can be found by minimizing the distance (defined as Equation 2 in \citet{LP2019}) between an isochrone and member stars.

The left panel of Figure \ref{fig:iso_fitting} displays the member candidates (red dots) and the initial sample (blue dots) distribution and its best fit isochrone (black solid curve) in the CAMD. The result shows that the age of the stellar ``snake" is around 33.8\,Myr, and the corresponding metallicity is around 0.027. But if we remove the member candidates with large color indexes, e.g., $G_{BP}-G_{RP}>3.0$, then the best fit age is 24.5\,Myr, and the corresponding metallicity is about 0.0152. The isochrone in this case is displayed with black dashed-dot curve for comparison. The spectroscopic data favour a larger metallicity, which has been discussed in Section \ref{sec:z_rv}. In this panel, we also illustrate two other isochrones with the age of 120\,Myr (green dashed) and 5\,Myr (green dashed-dot) which are used to remove the obvious outliers from the candidate members. 

We can use the technique of Turn-On \citep[i.e., TOn][]{Cignoni2010} to validate the age estimated from the isochrone fit. TOn is a transition point in the CMD of a very young stellar population, where the PMS joins the MS. Because the massive members have already entered in the stage of MS, a few field MS contaminators will lead to a challenge to distinguish the TOn point in the CAMD. 

In order to resolve the TOn point, we further purge the candidate members with more stringent criteria: (1) $(\mu_{l^*} - \bar{\mu}_{l^*,m})^2 + (\mu_{b} - \bar{\mu}_{b, m})^2<1.0\sigma_{\mu, m}^2$; (2)  keep only candidates with distances between 280\,pc to 340\,pc (see Figure \ref{fig:hist_dist_rv}); (3) $\rm RUWE<1.2$. The first two conditions can strictly constrain the members to be in the two cores (see the sub-panel (a) in Figure \ref{fig:data_distr}). In these two dense regions, the field contaminators are relatively few. The third condition makes sure the members have excellent astrometric solutions. Finally, we obtain a sub-sample of 634 candidate members with a very high confidence. Figure \ref{fig:iso_fitting} demonstrates the CAMD of the sub-sample in the right panel. From the insert sub-panel, one can easily see the TOn point, which is located at $G_{BP}-G_{RP}\sim1.0$\,mag where the stellar mass is $\sim$1.0\,$M_{\odot}$. It indicates that the age of this young population is around 30-40\,Myr. This result is well consistent with the age derived with the method of isochrone fit. 

%The age can also be validated with the $\log$g of part of members, listed in Table \ref{tab:stellar_para}. The average $\log$g (4.31$\pm$0.18) corresponds to a stellar age of 30-40\,Myr.

In estimating the age,  we have already determined the best fit isochrone. With this isochrone, stellar masses are then estimated. More than 80\% members still do not pass the TOn point and enter in the stage of MS, i.e., $G_{BP}-G_{RP}<1.0$\,mag. This suggests that more than 80\% members have masses smaller than 1.0\,$M_{\odot}$. The remaining $\sim$20\% members have masses larger than the solar mass, the maximum mass is $\sim$3.0\,$M_{\odot}$. From the best fit isochrone, it is easy to derive the present-day mass function of the tidal tail. By fitting the mass function with a series of initial mass function (IMF) of \citet{Kroupa2001}, a total mass of $\sim$2500\,$M_{\odot}$ is estimated for the stellar ``snake" (the blue curve is the best fitted mass function in the top panel of Figure \ref{fig:mass}). 

Using the color as a proxy of stellar mass, we try to check the theory of mass segregation \citep{BT1987}, which is usually followed by the classical tidal tails. As one expect, we do not find any strong evidence for mass segregation within 125\,pc from the cluster center. Beyond 170\,pc, however, there is a weak tendency that massive stars are expected to move toward the center of a system, while lower-mass stars can more easily evaporate from a cluster, as shown in the bottom panel of Figure \ref{fig:mass}. Note that the tendency of first rising then descending between 10 and 110\,pc is similar with what was found in the tidal tails of the Hyades discovered by \citet{MA2019}, who thought that mass segregation, remains unsuccessful, most likely because of the sensitivity limit for radial velocity measurements with {\it Gaia}. In this work, no strong evidence is found for mass segregation, probably because that the ``snake" actually is not a dynamically tidal tail, but a primordial morphology.
%Based on the best fitted isochrone, 

\subsection{The two open clusters}
In the sky region of our interest, we notice two interesting open clusters (as shown in Figure \ref{fig:data_distr}): one is the open cluster BBJ 1 (the cyan dots) discovered by B20, the other is an unknown open cluster (the black dots), which has not been reported so far \citep{Kharchenko2013,CG2018,LP2019,Sim2019,CG2020}, we temporarily name it as Tian 1 in this study. To compare the two clusters with the ``snake", we select the candidate members for the Tian 1 and BBJ 1 according to their coordinates, similar to what did for the above two cores. We roughly estimate ($l_0, b_0,\epsilon_0$) = ($213\degr, 8.0\degr, 1.5\degr$) for the Tian 1, and ($l_0, b_0,\epsilon_0$) = ($224\degr, -13.8\degr, 1.0\degr$) for the BBJ 1. Additionally, we remove the obvious outliers for both the clusters in the 2D proper motion space. Thus, we obtain 117 and 174 candidate members for the Tian 1 and BBJ 1, respectively. The bottom two sub-panels in Figure \ref{fig:hist_dist_rv} display the distance and radial velocity distributions of the Tian 1 (black) and BBJ 1 (cyan), respectively. The average distances (the radial velocities) are 466.7$\pm$16.7\,pc (15.1$\pm$5.9\,\kms) and 371.9$\pm$3.7\,pc ($24.8\pm3.7$\,\kms), for the two clusters, respectively.  The distance of BBJ 1 we measured are well consistent with the value ($\sim374$\,pc) obtained in B20. The distributions of the radial velocities look not so good, because there are only 8 and 23 available radial velocities for the two clusters. Unfortunately, no available metallicity is obtained for the open cluster BBJ 1.

The age of BBJ 1 is about 35\,Myr derived by B20. This age is well consistent with the ``snake", which indicates that they probably originate from the same population. The age of the Tian 1 is about 120\,Myr, which is simply estimated by the isochrone fitting (see the CMD for the Tian 1, shown by the black dots and black dashed curve in the left sub-panel of Figure \ref{fig:iso_fitting}).

\section{Discussion}\label{sec:dis}
In this section, we will simply analyze the possible mechanism of the ``snake" formation and then briefly discuss its potential values for the future study.%

With the population properties derived in Section \ref{sec:results}, we know that the stellar ``snake" has a long tail and two dissolving cores in the same population. To be the first choice, one easily considers that this is a structure of the tidal tail. According to theoretical and numerical studies, the so-called tidal tails are usually thought to form from star clusters. Due to the impacts of the Galactic gravitational tides, e.g., by passing molecular clouds, disk shocking, spiral arm passages, or other unknown events, star clusters will continuously lose members if members are no longer gravitationally bound. The tails consist of stars escaped from the cluster, which lead and/or trail their parent cluster along the orbit \citep{CR2006}. Under this prevailing theory of tidal tail formation and evolution, \citet{Kharchenko2009} carried out a set of high resolution N-body simulations to investigate how the shape parameters of open clusters vary with the time that the Galaxy's external force acts on star clusters with different initial masses, Galactocentric distances, and rotation velocities. The results tell us that the mass lose is only around 10\% of the initial masses whatever the initial conditions are (See their Figure 2). Their Figure 3 also clearly shows that at $t=50$\,Myr the star cluster is just stretched into an ellipsoid. Though the predicted picture can well match with the tidal tail of the Hyades with an age of several hundred Myr discovered by \citet{MA2019} and \citet{Roser2019}, it can not explain our stellar ``snake" with an age of only 30-40\,Myr (an order of magnitude younger than the ages of previously known tidal tails). Therefore, this kind of structures are generally considered to form primordially from a huge MC in the hierarchical models \citep[B20;][]{Kounkel2019,Kounkel2020}.

\begin{figure}[!t]
\centering
\includegraphics[width=0.45\textwidth, trim=0.1cm 0.07cm 0.cm 0.cm, clip]{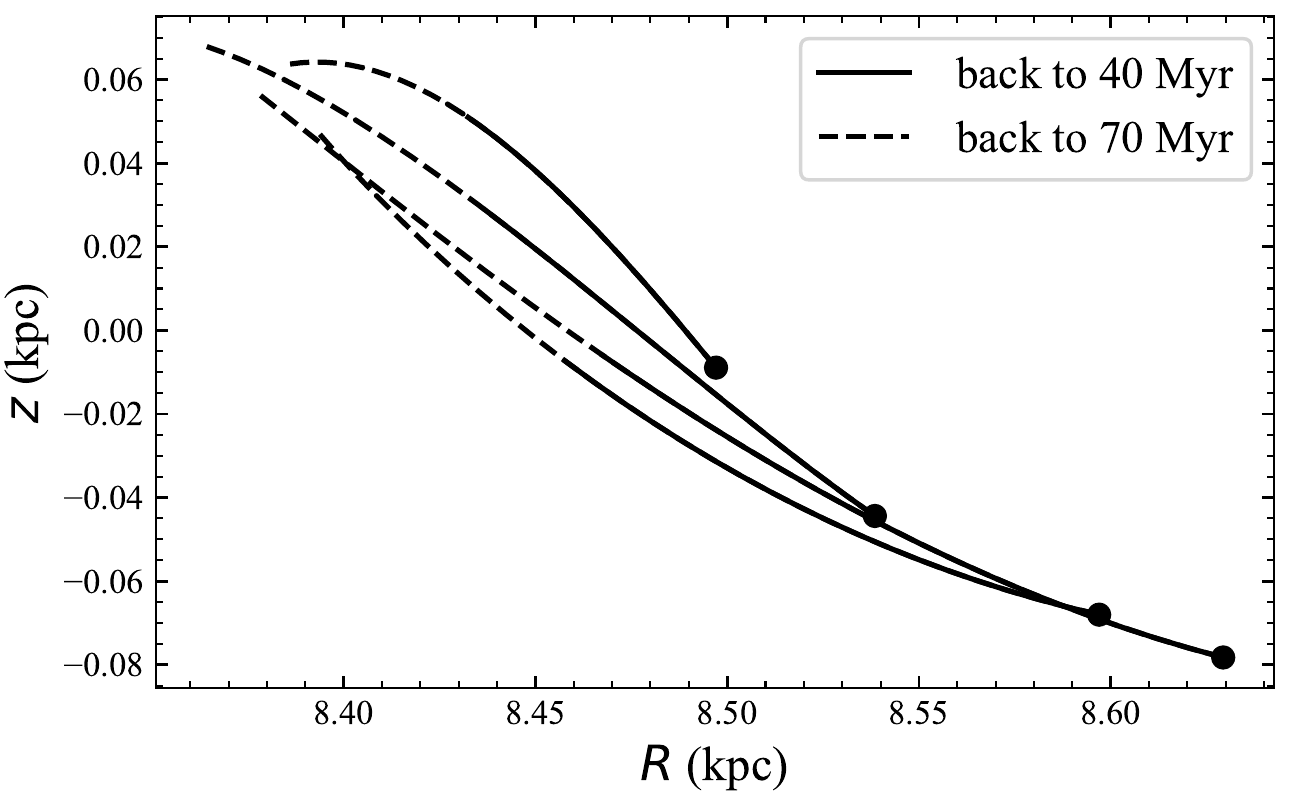}
\caption{ The orbits of four member stars in the R-z space. The four stars are selected from the different parts in the ``snake" structure. The four dots mark the present locations of the four stars, the solid curves represent their orbits integrated back to 40\,Myr, while the dashed curves display the orbits further integrated back to 70\,Myr. It seems that the fours stars ever stayed closer than the locations at the present. It indicates the ``snake" is probably expanding.
}\label{fig:orbits}
\end{figure}

We select four stars whose radial velocities are well measured (with errors $<$ 0.1\kms) by APOGEE from the two cores and the end part of the tail, respectively. The obits of the four stars can be solved accurately under the potential model of \citet{Bovy2015}. Figure \ref{fig:orbits} displays the four orbits, which are obtained by integrating back 40\,Myrs (the solid curves) and 70\,Myrs (the dashed curves) for the four stars. As the figure shown, the four stars seem to be stay closer before. It indicates that the ``snake"  is probably expanding. But this need more radial velocities with high precision for further validation. 

%the four obits roughly cross together. It indicates that the members are probably have the same origin. As for the complete orbits, a following work is preparing in detail.

 %presents a great challenge to the prevailing theory of tidal tail formation and evolution.
 %The shape of the tidal tail indicates that its parent cluster has probably experienced one or several strong disruptions during the course of its life. 
%Besides internal factors (self-gravitation and rotation), the shape of an open cluster mainly depends on the external forces, e.g. by the Galactic tidal field, which is theoretically thought to act in two steps: (1) stretching the cluster into an ellipsoid, and (2) producing cluster tails outpouring from the ellipsoids \citep{Wielen1974, Wielen1985}. 
%What mechanism stretches the parent cluster into a tail of size over 200\,pc in such a short time? 
% It suggests that this structure probably originates from a large open star cluster, which consists of a few thousand member stars.

 The new discovery is important at least in two aspects. On one hand, the ``snake" structure may be a good extension to the sky coverage and census of the Vela OB2 complex. As one known, the Vela OB2 region hosts a complex constellation of sub-populations with ages in the range 10 to 50\,Myr. The age of the ``snake" is well consistent with that of the open cluster BBJ 1, which is one of most important parts of the 260\,pc wide filamentary structure discovered by B20 in the Vela OB2 region. Additionally, the motion of the ``snake" is coherent with the structure of B20 in the two directions, i.e., $v_{r}$ and \mub, but the values of \mul\ are not. This perhaps is indued by some unknown perturbations. Moreover, the two structures are located in the coherent distances from us. The sky coverage of the ``snake" span at least in the three constellations, i.e., the Monoceros, Orion, and Taurus. If the ``snake" really formed in the same environment (e.g., a giant molecular cloud) with the structure of B20, the region of the Vela OB2 complex will be extended by a factor of $\sim$2. However, we need more observational information, e.g., the radial velocities, metallicities with high precision for more member stars of both the ``snake" and BBJ 1, to further confirm the relationship of the two structures. We are preparing for the follow-up observations in the LAMOST Medium-Resolution Spectroscopic Survey \citep[LAMOST-MRS,][]{LC2020}.

On the other hand, a number of member stars of the ``snake" are located in the Orion region, particularly in the $\lambda$ Ori (see Figure \ref{fig:data_distr}). The Orion complex is thought to form part of the Gould Belt \citep{poppel1997}, which is a large ($\sim$1\,kpc), young (30 - 40\,Myr), and ring-like structure tilted $\sim20\degr$ to the Galactic plane discovered by Herschel in 1847. For a long time, this is not consistent with the age measured in the Orion complex. The oldest part in the Orion complex was believed to be no more than 13\,Myr old. Until the early of 2019, \citet{Kos2019} extended the oldest age to 21\,Myr. However, there still exists a gap between the Gould Belt and the Orion complex in the history of star formation. Remarkably, the age of the stellar ``snake" well agrees with the history of Gould Belt. However, it is worth to mention that the existence of the Gould Belt is challenged by a recent structure -- Radcliffe Wave, i.e., a Galactic-scale gas wave in the solar neighborhood, discovered by \citet{Alves2020}. This finding is inconsistent with the notion that the clouds are part of a ring, disputing the Gould Belt model. So the stellar ``snake" probably bridges the gap between star formation near the region of the Orion complex and the history of Gould Belt's formation (if the Belt exists). The stellar ``snake" is perhaps a fine structure on top of the Radcliffe Wave.

%where exists plentiful star forming regions . 
%of this stellar population is only 
%This In this section, we will briefly discuss the importance of the newly discovered tidal tail in two aspects. %and also will talk about 
% of tidal tail formation and evolution

\section{Conclusion} \label{sec:conclusion}
A young snake-like structure (dubbed stellar ``snake") in the solar neighborhood is found from the {\it Gaia} DR2. Using the FOF algorithm, we find 1989 candidate members from the stellar ``snake". Its average distance is about 310\,pc from the Sun. The length (x direction) and width (y direction) are over 200\,pc, but thickness (z direction) is only about 80\,pc. %The structure is not typical of double-armed tidal tails.  
%This structure does not present a leading tail, and has only a long trailing tail. 
Interestingly, the stellar ``snake" includes two dissolving cores in its head. The two cores can be clearly distinguished in the 6D phase space, which probably form primordially from a huge MC with two sub-groups. Using the techniques of the isochrone fitting and TOn, we measure the age of the stellar population. Surprisingly, the age of this population is so young (only 30-40\,Myr) that its formation can not be explained with the theory of tidal tails. It suggests that this filamentary structure may be hierarchically primordial, rather than the result of tidal stripping or dynamical processing, even though the ``snake" is probably expanding. This also could be deduced by the fact that no any strong evidence can be found for the theory of mass segregation \citep{BT1987}, which is usually followed by the classical tidal tails. By fitting the mass function with a series of IMF of \citet{Kroupa2001}, we estimate a total mass of $\sim2500M_{\odot}$ for the stellar ``snake".

The finding may be a good extension to the sky coverage and census of the Vela OB2 complex. According to the 5D phase information and the age, we suspect that the ``snake" and the open cluster BBJ 1 were born in the same environment. If so, the ``snake" could extend the region of the Vela OB2 to another three constellations, i.e., the Monoceros, Orion, and Taurus, and supplement more sub-structures for the census of the Vela OB2 complex. This finding is useful to understand the history of the Vela OB2 formation and evolution, and provide some clues to connect the complex with the Galactic-scale structures, such as the Gould Belt and the Radcliffe Wave  \citep{Alves2020}. The age of the ``snake" well matches with that of the Gould Belt. In the sky region of our interest, we detect one new open cluster, which is named as Tian 1 in this work.

%prevailing




 \acknowledgements
%The author thanks Di Li  for his very constructive comments,
H.-J.T. thanks Min Fang, Jiaming Liu, Hao Tian, Lei Liu, Boquan Chen, Chao Liu, Di Li, Hongsheng Zhao and Yougang Wang for the helpful discussions, and thanks Hongsheng Zhao for revising the whole manuscript, also acknowledges the National Natural Science Foundation of China (NSFC) under grants 11873034, and the Cultivation Project for LAMOST Scientific Payoff and Research Achievement of CAMS-CAS. This work has made use of data from the European Space Agency (ESA) mission
{\it Gaia} (\url{https://www.cosmos.esa.int/gaia}), processed by the {\it Gaia}
Data Processing and Analysis Consortium (DPAC, \url{https://www.cosmos.esa.int/web/gaia/dpac/consortium}). Funding for the DPAC
has been provided by national institutions, in particular the institutions
participating in the {\it Gaia} Multilateral Agreement.
The Guo Shou Jing Telescope (the Large Sky Area Multi-Object Fiber Spectroscopic Telescope, LAMOST) is a National Major Scientific Project built by the Chinese Academy of Sciences. Funding for the project has been provided by the National Development and Reform Commission. LAMOST is operated and managed by National Astronomical Observatories, Chinese Academy of Sciences.



\end{document}